\documentclass[twocolumn,tighten]{aastex63}
\usepackage{mathtools}
\usepackage{txfonts} 
\usepackage{hyperref}
\usepackage{color}

\defcitealias{PaperI}{EHTC~I}
\defcitealias{PaperII}{EHTC~II}
\defcitealias{PaperIII}{EHTC~III}
\defcitealias{PaperIV}{EHTC~IV}
\defcitealias{PaperV}{EHTC~V}
\defcitealias{PaperVI}{EHTC~VI}
\defcitealias{PaperVII}{EHTC~VII}
\defcitealias{PaperVIII}{EHTC~VIII}
\defcitealias{PaperIX}{EHTC~IX}
\defcitealias{Narayan_2021}{N21}
\defcitealias{Gelles_2021}{G21}

\turnoffeditone

\newcommand{\m}{M87*}
\newcommand{\s}{Sgr\,A*}
\begin{document}                                              

\title{Multifrequency Analysis of Favored Models for the Messier 87* Accretion Flow
}
\shorttitle{Multifrequency Analysis of Favored M87* Simulations}

\author[0000-0002-7179-3816]{Daniel~C.~M.~Palumbo}
\affil{Center for Astrophysics $\vert$ Harvard \& Smithsonian, 60 Garden Street, Cambridge, MA 02138, USA}
\affiliation{Black Hole Initiative at Harvard University, 20 Garden Street, Cambridge, MA 02138, USA}

\author[0000-0002-5518-2812]{Michi~Baub\"ock}
\affil{Department of Physics, University of Illinois at Urbana-Champaign, 1002 West Green Street, Urbana, IL, 61801, USA}

\author[0000-0001-7451-8935]{Charles~F.~Gammie}
\affil{Department of Astronomy, Department of Physics, and NCSA, University of Illinois at Urbana-Champaign, 1002 West Green Street, Urbana, IL, 61801, USA}
%Analyzing these images requires comparison to time-dependent general relativistic magnetohydrodynamic (GRMHD) simulations imaged with general relativistic ray tracing (GRRT). These steps form a numerical modeling pipeline with computational and physical limitations on accuracy. 
    
\begin{abstract}
    The polarized images of the supermassive black hole Messier 87* (M87*) produced by the Event Horizon Telescope (EHT) provide a direct view of the near-horizon emission from a black hole accretion and jet system.  The EHT theoretical analysis of the polarized M87* images compared thousands of snapshots from numerical models with a variety of spins, magnetization states, viewing inclinations, and electron energy distributions, and found a small subset consistent with the observed image. In this article, we examine two models favored by EHT analyses: a magnetically arrested disk with moderate retrograde spin and a magnetically arrested disk with high prograde spin. Both have electron distribution functions which lead to strong depolarization by cold electrons. We ray trace five snapshots from each model at 22, 43, 86, 230, 345, and 690 GHz to forecast future VLBI observations and examine limitations in numerical models. We find that even at low frequencies where optical and Faraday rotation depths are large, approximately rotationally symmetric polarization persists, suggesting that shallow depths dominate the polarization signal. However, morphology and spectra suggest that the assumed thermal electron distribution is not adequate to describe emission from the jet. We find 86 GHz images show a ring-like shape determined by a combination of plasma and spacetime imprints, smaller in diameter than recent results from the Global mm-VLBI array. We find that the photon ring becomes more apparent with increasing frequency, and is more apparent in the retrograde model, leading to large differences between models in asymmetry and polarization structure. 
\end{abstract}

\section{Introduction}

In 2017, the Event Horizon Telescope (EHT) observed the supermassive black hole at the heart of the supergiant elliptical galaxy Messier 87. The first set of EHT results analyzed only total intensity (Stokes $I$) images of this black hole (hereafter \m{}), and found a few striking results in light of a library of synthetic images derived from general relativistic ray tracing (GRRT) of general relativistic magnetohydrodynamical (GRMHD) simulations \citep[][hereafter EHTC I-VI]{PaperI, PaperII, PaperIII, PaperIV, PaperV, PaperVI}. First, the image library produced in \citetalias{PaperV} yielded a calibration from the apparent size of observed, blurry ring-like images to black hole images with known masses, enabling the first horizon-scale measurement of a SMBH mass. Second, although most models were inconsistent with non-EHT constraints such as jet power, the vast majority of models were capable of reproducing the image morphology at the 230 GHz frequency (and resolution) of the EHT. A similar morphology was recently recovered from data taken the following year \citep{EHTM87_2018}.

\begin{figure*}[t!]
    \centering
    \includegraphics[width=0.98\textwidth]{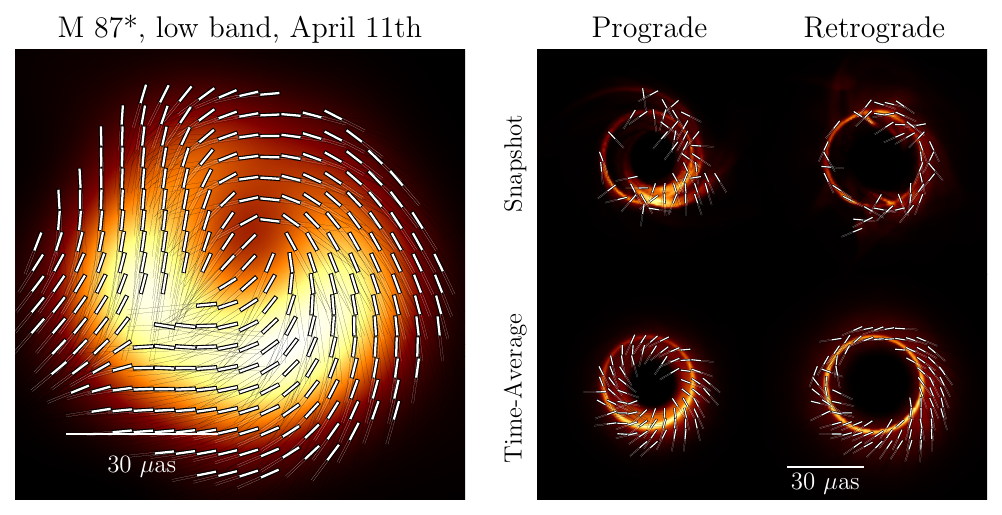}
    \caption{Comparison of single snapshots and time-averaged 230 GHz images of the two fiducial models to the \m{} image. Here and elsewhere in the paper, simulation images have been rotated to point the approaching jet at $288^\circ$ east of north \citep[see, e.g.][]{Walker_2018}, and white ticks  indicate the electric vector position angle wherever the fractional polarization exceeds 1\% and the total intensity exceeds 10\% of its peak value.}
    \label{fig:m87}
\end{figure*}

In contrast to the total intensity results from the 2017 data, the analysis of linearly polarized images of \m{} from the same observing campaign was highly model-constraining \citep[][hereafter EHTC VII-VIII]{PaperVII, PaperVIII}. The polarized image morphology indicated a coherent polarization spiral and low overall polarization. The EHT analysis defined a number of image-domain observables for comparison against GRMHD, metrics by which only fifteen models pass in \citetalias{PaperVIII}. The vast majority of passing model snapshots came from Magnetically Arrested Disks (MADs), accretion flows in which the magnetic pressure in the inner accretion flow is dynamically important and sufficient to balance the inward ram pressure of the accreting plasma \citep{Ichimaru1977,Igumenschchev_2003,Narayan2003}. The low polarization of \m{} favored models with higher values of $R_{\rm low}$ and $R_{\rm high}$, parameters introduced by \citet{Mosci_2016} to tune the electron temperature with respect to the ion temperature. When accretion rates are adjusted to hold the average flux constant at an observed value, higher values of these parameters indicate colder electrons, which produce stronger Faraday effects in the flow, in turn depolarizing the image. More recently, circular polarization was analyzed in the same dataset in \citet{PaperIX}, hereafter \citetalias{PaperIX}. Though the morphology of circular polarization was not well-constrained by EHT data, upper bounds on the circular polarization magnitude yielded a further preference for magnetically arrested disks with cold electrons.

The restriction of viable GRMHD model parameters to a small subset invites deep, rather than broad, explorations of observable properties for future observing campaigns and for observations at other frequencies. Indeed, \citetalias{PaperVIII} examined the 230 and 345 GHz polarized image properties and variability of several passing models in its forecast of future observations. Most excitingly, recent observations of \m{} at 3 mm by the Global mm-VLBI array (GMVA) indicate a structural connection between the accretion flow and jet in the form of a bright ring with a central flux depression \citep{Lu2023}. 

In this article, we extend the multifrequency examination of two favored models for \m{} in light of growing considerations of next-generation expansion of the Event Horizon Telescope. We probe image quantities of interest at 22, 43, 86, 230, 345, and 690 GHz frequencies; at lower frequencies, images are dominated by jet emission, while at higher frequencies (particularly 690 GHz), image morphology is dominated by lensed emission from the photon ring, the sharp ring of light corresponding to photons which half-orbit the black hole at least once before reaching the observer \citep[see, e.g.,][]{Johnson_2020}. In examining frequencies far above and below 230 GHz, we push the limits of the numerical models; we thus also use this analysis as an opportunity to highlight priorities in future methodological improvements to GRMHD and GRRT that will be necessary for interpretation of ground-based very long baseline interferometry (VLBI) of \m{}. \edit1{Concretely, we seek to find where on the low frequency end these simulations break down, and what behavior to expect at high frequencies as VLBI advances enable sub-mm observations at 345 GHz and above. We particularly examine image-domain intrinsic and morphological properties that are illustrative of spacetime features such as the photon ring.}

In \autoref{sec:models}, we introduce the two favored \m{} models. In \autoref{sec:quantities}, we compute total intensity and polarized image quantities for the models across six frequencies. We conclude with a discussion of simulation development and observational priorities in \autoref{sec:discussion}.

\section{Favored M~87* Models}
\label{sec:models}

\begin{figure*}
    \centering
    \includegraphics[width=\textwidth]{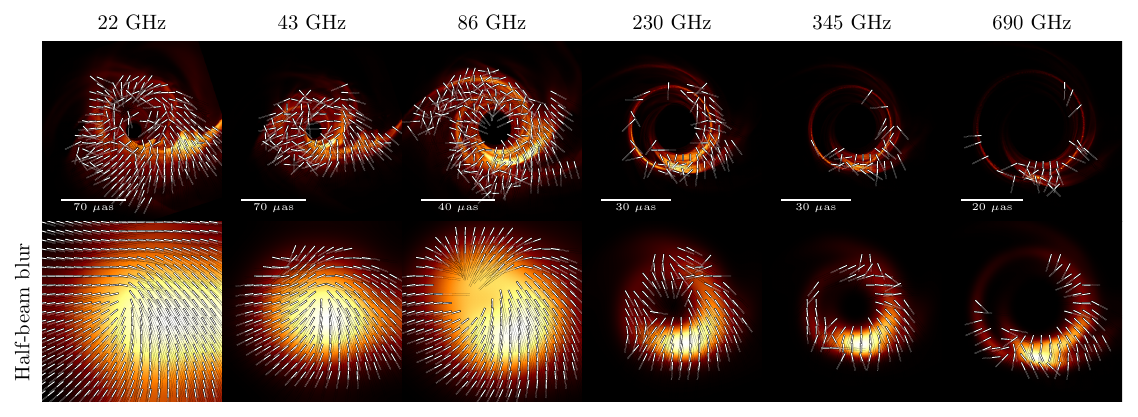}
    \caption{Snapshots of the $a_*=+0.94$ MAD with $R_{\rm low}=10$, $R_{\rm high}=80$. Bottom row: the same snapshots blurred by half of the diffraction-limited beam for an Earth-diameter baseline. Note the decreasing field of view with increasing frequency. }
    \label{fig:pro_ims}
\end{figure*}

\begin{figure*}
    \centering
    \includegraphics[width=\textwidth]{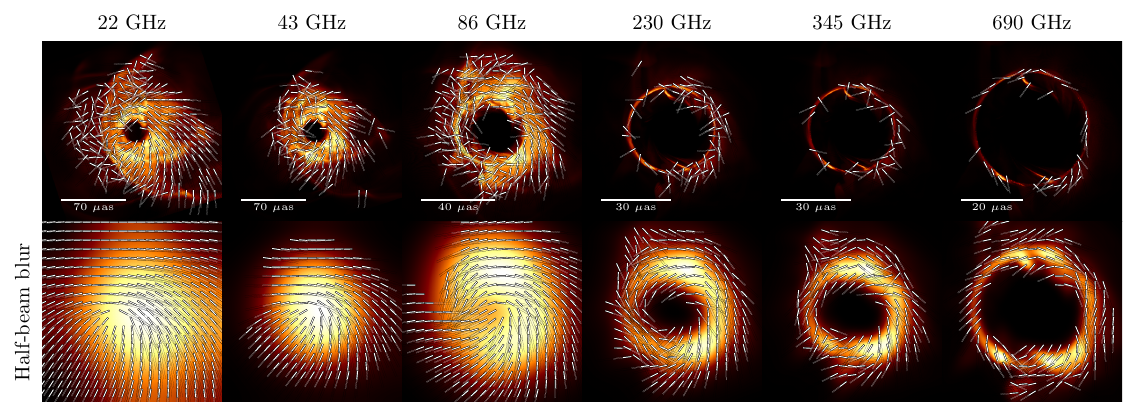}
    \caption{Same as \autoref{fig:pro_ims}, but for the  $a_*=-0.5$ MAD with $R_{\rm low}=1$, $R_{\rm high}=160$. }
    \label{fig:ret_ims}
\end{figure*}

% The theoretical analyses of the 2017 EHT data on \m{} analyzed a large library of GRMHD simulations, comparing thousands of snapshots to the observed total inten polarized images in \citetalias{PaperVII}.

Here, we examine GRMHD simulations used for comparison to EHT data throughout the total intensity, linear polarization, and circular polarization analyses carried out by the EHT. The models we will examine were produced using \texttt{iharm3D} \citep{Gammie_HARM_2003, IHARM3d_prather} and ray traced with \texttt{ipole} \citep{IPOLE_2018}. Additional details on the image generation process may be found in \citet{Wong_2022}.

We closely examine two models which pass the observed image comparison criteria and jet power constraints for \m{} (see \citetalias{PaperV}, \citetalias{PaperVIII}, and \citetalias{PaperIX}): both are MADs, one with dimensionless spin $a_*=+0.94$, $R_{\rm low}=10$, and $R_{\rm high}=80$, and another with $a_*=-0.5$, $R_{\rm low}=1$, and $R_{\rm high}=160$.  As defined in \citet{Mosci_2016}, the parameters $R_{\rm low}$ and $R_{\rm high}$ set the relative temperatures of ions ($T_i$) and electrons ($T_e$) according to
\begin{align}
    \frac{T_i}{T_e} &= R_{\rm low}\frac{1}{1+\beta^2} + R_{\rm high} \frac{\beta^2}{1+\beta^2}.
\end{align}
Here, $\beta = P_{\rm gas}/P_{\rm mag}$ is the ratio between the gas and magnetic pressure in a given GRMHD cell.  \edit1{This prescription is motivated by models for heating in a collisionless plasma \citep[e.g.][]{Quataert_Gruzinov_1999, Howes_2010, Kawazura_2017} , where the ratio of ion to electron heating increases as $\beta$ increases.  The $R_{\rm high}$ prescription models the outcome of this process in an expression that is suitable for radiatively post-processing GRMHD data. In this way, this heating prescription tunes temperature distinctly between the relatively strongly magnetized jet and weakly magnetized disk.}

\edit1{In these models, a non-zero black hole spin is expected to impact the apparent rotation of material near the horizon, winding up matter trajectories and magnetic fields near and within the ergosphere. Previous work has shown that frame dragging often manifests visually in MADs as a twisting of polarization around the black hole shadow \citep{PWP_2020, Ricarte_2022, Emami_2023}. Spin is also expected to impact matter orbits even in these models, where the innermost stable circular orbit does not play a significant role \citep{Conroy_2023}.} In this model set, a negative sign of spin refers to a black hole that is spinning retrograde with respect to its large scale accretion flow. We select these two parameter sets in order to have two spin magnitudes and both a retrograde and prograde model represented in our analysis. Note that we consider only a single global magnetic field polarity.  

\autoref{fig:m87} compares a snapshot and the time-averaged image of these models ray traced at 230 GHz to the low band survey-averaged reconstruction of \m{} from \citetalias{PaperVII}. In each case, we use the angular gravitational radius (equivalently, mass-to-distance ratio) of 3.62 $\mu$as  used in \citetalias{PaperV} and \citetalias{PaperVIII} simulations based on stellar absorption line data \citep{Gebhardt_2011} which is approximately 5\% lower than is inferred by the EHT. As discussed extensively in \citetalias{PaperVIII}, even if the numerical models are a perfect representation of the \m{} accretion flow, it is improbable that any of the individual synthetic images will be consistent with the data to within thermal errors since the model is chaotic, and the assumptions of the simulation do not exactly mimic nature.  Instead we expect the distribution of quantities measured on the data to support some simulations more than others. Thus, model libraries for image comparison help to bound the phenomenological space of accretion flows.

In advance of new observations of \m{} at 230 GHz and other frequencies, we take five snapshots of each model, sampled every $1000 G M/c^3 \simeq 383$ days starting after 26000 $G M/c^3$ have elapsed in the fluid simulation, and ray trace them at 6 frequencies. The first such snapshot of each model is shown before and after blurring by half of the nominal Earth-diameter interferometric resolution in \autoref{fig:pro_ims} and \autoref{fig:ret_ims}, where the nominal resolution is $\lambda/(2 R_\oplus)$. This half-beam resolution is $\sim10$ $\mu$as at 230 GHz and scales linearly with wavelength. We choose to use half of the beam because this level of super-resolution is typical in modern Bayesian imaging methods and it is near the limit reached with well-sampled coverage (see the discussion of resolution in \citetalias{PaperIV}, \citealt{Palumbo_2019} and \citealt{SgrA_PaperIII}).

As there is a degree of freedom in the mass accretion rate in the simulations, we may scale the accretion rate to match the observed flux at a single wavelength. As in \citetalias{PaperV}, we normalize the accretion such that the average flux of the five selected snapshots at 230~GHz is 0.5~Jy.  Notice that although the observed compact flux is consistent with 0.5~Jy it is not known precisely and could be as large as $\sim 1$ Jy (see also the discussion of the compact flux constraints in \citetalias{PaperIV}).

Before any quantitative analysis, a few features are immediately apparent. In unblurred images, the ``inner shadow'' feature corresponding to the lensed appearance of the near side of the event horizon is present \citep{Chael_2021}. At low frequencies, the first (``$n=1$'') photon ring sub-image disappears, indicative of large optical depths. However, in the retrograde model, the photon ring is still visible at 86\,GHz. Even at low frequencies where the total optical and Faraday depths encountered by light rays are generally large, polarization appears ordered even before blurring, suggesting that magnetic fields are ordered throughout the flow and that the observed EVPA is dominated by shallow Faraday rotation depths for which polarization is preserved on the path to the observer in the absence of an external screen. Even at half of the Earth-diameter resolution, 22 GHz and 43 GHz images do not show ring-like structures, while 86 GHz images show a slight central brightness depression.

% \begin{enumerate}
%     \item In unblurred images, the ``inner shadow'' feature corresponding to the lensed appearance of the near side of the event horizon is present \citep{Chael_2021}.
%     \item At low frequencies, the first (``$n=1$'') photon ring sub-image disappears, indicative of large optical depths. However, only in the retrograde model, the photon ring is still visible at 86 GHz.
%     \item Even at low frequencies where the total optical and Faraday depths encountered by light rays are generally large, polarization appears ordered even before blurring, suggesting that magnetic fields are ordered throughout the flow and that the observed EVPA is dominated by shallow Faraday rotation depths for which polarization is preserved on the path to the observer in the absence of an external screen.
%     \item Even at half of the Earth-diameter resolution, 22 GHz and 43 GHz images do not show ring-like structures, while 86 GHz images show a slight central brightness depression.
% \end{enumerate}

\section{Image Quantities}
\label{sec:quantities}

\begin{figure*}
    \centering
    \includegraphics[width=0.98\textwidth]{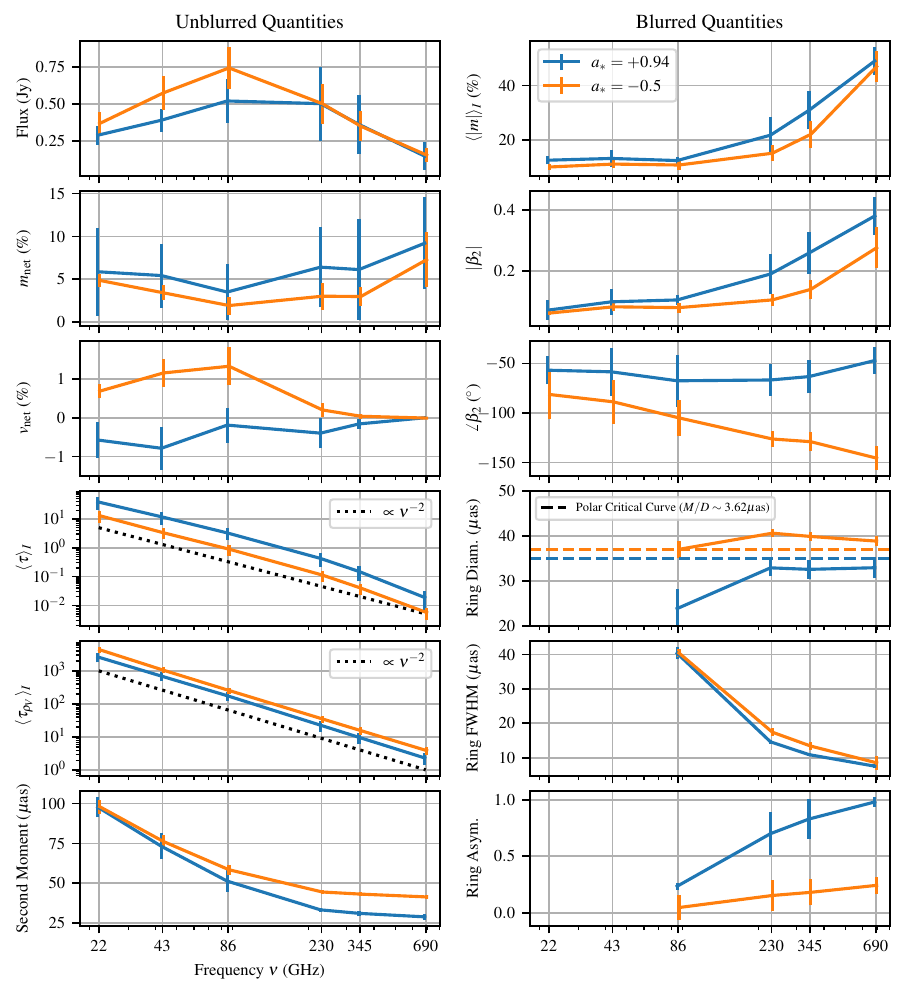}
    \caption{Image quantities enumerated in \autoref{sec:quantities} for the two fiducial \m{} models. The left column contains quantities computed from the unblurred image, while the right column contains blurred image quantities. At 22 and 43 GHz, the morphology of the image blurred to half of the nominal beam is not ring-like, so no ring is fit. Small horizontal offsets at a single frequency are for visual clarity only.}
    \label{fig:quantities}
\end{figure*}

We now compute 12 quantities of interest in both total intensity and polarization. The first six are computed on the unblurred image:
\begin{enumerate}
    \item The total image flux in janskys (Jy). 
    \item The unresolved fractional linear polarization $m_{\rm net}$, equivalent to the magnitude of the total linear polarization (a complex scalar) divided by the total Stokes $I$ flux.
    \item The unresolved fractional circular polarization $v_{\rm net}$, equivalent to the total Stokes $V$ flux divided by the total Stokes $I$ flux.
    \item The Stokes $I$ pixel-weighted average optical depth $\langle\tau\rangle_I$, computed along rays through the full volume of each fluid frame.
    \item The Stokes $I$ pixel-weighted average Faraday rotation depth $\langle\tau_{\rho_V}\rangle_I$.
    \item The image second moment, measured by diagonalizing the image covariance matrix, corresponding to the Gaussian size measured by short baselines that do not resolve the source structure (see \citet{Issaoun_2019} for details). For ring-like images, this corresponds approximately to the ring diameter at the 10\% level.
\end{enumerate}

The latter six quantities are computed after blurring the image by half of the nominal beam corresponding to Earth-diameter baselines at each frequency. The blurred quantities are as follows:
\begin{enumerate}
    \item The average resolved (pixel-wise) linear polarization fraction $\langle | m | \rangle_I$.
    \item The amplitude of the rotationally symmetric polarization spiral mode $\beta_2$, defined in \citep{PWP_2020}. For nearly rotationally symmetric polarized images in which the $\beta_2$ mode is dominant, $|\beta_2|$ is close to $\langle | m | \rangle_I$.
    \item The phase of $\beta_2$, $\angle\beta_2$, which encodes the handedness and rotationally symmetric orientation of the electric vector position angle (EVPA) spiral. In \citetalias{PaperVIII}, this metric was the most discriminating of accretion parameters.
    \item The diameter $D$ of a fitted asymmetric ring, fit using \texttt{VIDA} \citep{VIDA}. Notice that this is not meant to match the radius of the critical curve, but instead characterizes the radius of a thick geometric ring on the sky fit to the image.
    \item The thickness (full width at half of maximum, or FWHM) $w$ of the fitted ring.
    \item The asymmetry $A_1$ of the fitted ring, corresponding to the $m=1$ m-ring mode amplitude as defined in section 5.2.2 of \citet{SgrA_PaperIV}.
\end{enumerate}

The ring we fit in \texttt{VIDA} is analogous to that used in the image domain feature extraction in \citet{SgrA_PaperIV}. We use a third-order m-ring as specified by \texttt{VIDA}'s \texttt{CosineRing} template; we do not include a flux floor. \autoref{fig:quantities} shows all twelve quantities as a function of frequency for the two fiducial models, with error bars indicating one standard deviation over the five ray traced snapshots at each frequency. 

In general, our results suggest promising connections between low frequency images and signatures of the black hole spacetime. In both models, the total image flux rises between 22 and 86 GHz, with a particularly dim 22 GHz flux. Together with the small $\sim 200\mu$as angular extent of the 22 GHz image, this suggests a dim jet that is contrary to observations \citep{Hada_2017, Kim_2018,Walker_2018}, and likely reflects limitations of the electron distribution function post-processing in jet-dominated images, and a tendency of the GRRT to underproduce large scale flux, particularly when large fractions of the jet volume are omitted due to limited simulation volume and low particle densities (see section 7.3 of \citetalias{PaperV} for a discussion of GRMHD floors). 

Meanwhile, the prograde and retrograde model have opposite signs of unresolved circular polarization, each of which is stable over time and frequency. Given that we consider only a single overall magnetic field polarity, the absolute sign cannot be straightforwardly interpreted between the two models (see \citetalias{PaperIX} for a discussion of field polarity). However, as discussed in detail in \citet{Mosci_2021} and \citet{Ricarte_2021}, Stokes $V$ is often stable in sign over time in near face-on MADs, as the overall magnetic field geometry changes little in traditional torus-fed GRMHD. The optical and Faraday rotation depths both fall approximately as $\nu^{-2}$ as expected, leading to a corresponding increase with frequency in all fractional linear polarization magnitudes ($m_{\rm net}$, $\langle | m | \rangle_I$, and $|\beta_2|$). 

The unblurred image second moment decreases with frequency from $\sim100$ $\mu$as to slowly approach the angular size of the photon ring. Notably, the effective full size of the unblurred image, twice the second moment, only exceeds the nominal beam size corresponding to an Earth-diameter baseline at 22 GHz (a beam size of $\sim210\,\mu$as as opposed to a doubled second moment of $\sim220\,\mu$as). This suggests that for all frequencies at or above 43 GHz, the \m{} inner accretion flow will be marginally resolved (sufficient to measure beam-scale morphological properties and brightness temperature of the core). However, the polarization is ordered even when typical optical and Faraday depths greatly exceed unity, including at 22 GHz. 

The spiral phase $\angle \beta_2$ is always negative, suggestive of clockwise apparent motion at all frequencies, with more radial EVPA at lower frequencies indicative of toroidal fields in the jet. The retrograde model trends towards more circular EVPA at higher frequencies, suggestive of more radial velocities and magnetic fields in the emitting region. However, this change with respect to frequency likely arises primarily from contributions from the direct image and the photon ring adding to produce an intermediate phase value. As found in \citet{Palumbo_2023_b2vis}, at 230 GHz and above, the spiral phase of the direct image and the first lensed image are each individually stable with respect to frequency. 

For images at 22 and 43 GHz, there is no apparent ring structure after blurring by half the nominal resolution, whereas at 86 GHz, the ring diameter is comparable to the FWHM of the ring thickness. As shown in the 86 GHz images in \autoref{fig:pro_ims} and \autoref{fig:ret_ims}, this ring structure has a very shallow central depression, resembling the larger depression observed at the core by \citet{Lu2023}. 

For images at 230 GHz and higher, the fitted ring diameter gradually approaches the photon ring size as frequency increases; dashed lines show the diameter of the critical curve (again with $M/D \sim 3.62 \mu$as) on an exactly polar observer's screen for the spin of each model \citep[see, e.g.][]{Gralla_2020_lensing}; these values approximate the warped critical curve size for small inclinations. At all frequencies, the morphology of the direct image of the plasma displaces the measured diameter from the critical curve; at higher frequencies, the relative contribution from the photon ring increases. 

The first-order ring asymmetry grows with frequency for both models, but is much larger in the prograde model, consistent with greater apparent azimuthal velocities when the large-scale angular momentum matches the black hole spin direction.

Due to our limited temporal sampling, we cannot make any detailed claims about time variability; however, our sampling suggests narrow distributions in most physical parameters with more significant variation in ring morphology over time.

\section{Discussion}
\label{sec:discussion}

We have closely examined two GRMHD models of the \m{} accretion flow favored by existing EHT analyses. One of the two models contained a black hole with spin oriented retrograde to the large scale accretion flow; both models contained fairly cold electrons relative to most models in the \citetalias{PaperV}, \citetalias{PaperVIII}, and \citetalias{PaperIX}  libraries. We ray traced these models at five times with six frequencies. For each image, we computed physical and observable quantities relevant to EHT observation and interpretation in order to probe limitations of the models and set expectations for future observing campaigns at these frequencies. \edit1{We found that several morphological quantities in both total intensity and linear polarization show strikingly different frequency evolution, providing key probes on black hole parameters, accretion history and jet-launching mechanisms at play in \m{} by adding sensitivity to black hole spin and its alignment with the large-scale accretion disk.}

While most quantities followed intuitive trends over frequency consistent with decreasing optical and Faraday rotation depths, this analysis indicated clear limitations of the GRMHD snapshot analysis pipeline, particularly in GRRT. Most apparent is the dim jet in all models, especially in lower frequency images. This strongly suggests that the existing electron distribution assignment scheme (the $R_{\rm low}$, $R_{\rm high}$ model) is not adequate to describe emission from the jet.  One possibility is that a more complicated scheme, where the electron temperature depends on both the plasma's gas to magnetic pressure ratio number $\beta$ and the plasma magnetization number $\sigma$, is needed.  More likely, in our opinion, is that a nonthermal tail on the distribution is required. In the jet the field strength and temperature, and therefore the characteristic thermal synchrotron frequency, decline rapidly with distance from the black hole.  Jet emission at 230 and 86 GHz is thus
likely dominated by emission from electrons with energy well above the thermal core of the distribution function, and is therefore sensitive to the presence of even a small nonthermal population of electrons.

One critical prediction of the model images is that the structure appears approximately ring-like at all frequencies with a shifting characteristic size (measured by the second moment) and peak brightness. This suggests that low frequency measurements (in this case, 22, 43, and 86 GHz) of sufficient resolution will see ring-like structures with radius only weakly related to spacetime imprints. It bears emphasizing that the close correspondence between the 230 GHz emission size and the photon ring diameter is a prediction that depends on the plasma state
and not solely on the spacetime geometry. Some models considered in \citetalias{PaperV} and disfavored by the later polarimetric analysis do not have as close a correspondence in apparent sizes.

Further, if the consistently visible inner shadow persists at low frequencies, space-based VLBI (such as 22 GHz Radioastron observations, discussed in \citealt{Bruni_2020}) may have sufficient resolution to see the inner shadow even if the flow is optically thick. Recent observations provide a new handle on this problem, as the 86 GHz images of \m{} produced by \citet{Lu2023} show a flux depression that could represent the blurred appearance of the inner shadow within an overall larger emission region than is present at 230 GHz and above. On larger scales, these observations show a central spine, which these simulations cannot reproduce due to the limited volume of the simulations (which is chosen for studies of the inner accretion flow in typical EHT work). 

These 86 GHz observations thus provide an interesting comparison for the images of the models we consider. 86 GHz appears to be a turning point in the impact of radiative transfer effects on the emission region with distinct behavior in the prograde and retrograde cases, especially given that the retrograde 86 GHz image includes the sharp photon ring (which is dominated by emission near the horizon). Though these simulations are known to under-produce jet emission, we note that the large angular size of the compact ring observed by \citet{Lu2023} ($64^{+4}_{-8} \mu$as) is more closely approached by the retrograde model, whereas the prograde emission geometry is more compact.  However, neither model produces a ring diameter consistent with the \citet{Lu2023} image.

As discussed first in \citetalias{PaperV} and \citetalias{PaperVIII}, among favored MAD simulations, retrograde models tend to produce emission at larger radii (corresponding to a lower inferred mass from the EHT image), along with lower Faraday rotation and optical depths than their prograde counterparts. Taken in concert with recent lower dynamical mass estimates for \m{} \citep{Liepold_2023}, the \citet{Lu2023} observations and GRMHD simulations examined here suggest that the 2017 EHT image of \m{} had significant contributions from emission outside the innermost few gravitational radii. In addition, the 86 GHz image suggests that even retrograde models do not produce large enough characteristic emission radii without modifications to the electron distribution function.

At high frequencies, the morphological quantities we analyzed showed a divergence between the prograde and retrograde model, particularly in the observed ring asymmetry and polarization spiral phase. These quantities can each vary greatly over time, but our analysis suggests that instantaneous frequency evolution can break degeneracies between prograde and retrograde flows. Thus, the proposed capability of the EHT to observe simultaneously at 230 and 345 GHz \citep{Doeleman_2023} may be critical to distinguishing accretion conditions.
%The EHT polarized analysis of \m{} found models capable of producing snapshots similar to the observed image; for \s{}, the additional variability observable broke model consistency, as no GRMHD model could reproduce the observed dynamics. It is likely that advances in GRMHD that address variability could alter other observed properties as well; it is thus critical that new models are tested against all published EHT images as they are released.

\acknowledgments{We thank our internal referees for their thoughtful comments. We thank Paul Tiede for his assistance in the application of \texttt{VIDA}. This work was supported by the Black Hole Initiative, which is funded by grants from the John Templeton Foundation (Grant 62286) and the Gordon and Betty Moore Foundation (Grant GBMF-8273) - although the opinions expressed in this work are those of the author(s) and do not necessarily reflect the views of these Foundations. This research used resources of the Oak Ridge Leadership Computing Facility at the Oak Ridge National Laboratory, which is supported by the Office of Science of the U.S. Department of Energy under Contract No. DE-AC05-00OR22725.  This research used resources of the Argonne Leadership Computing Facility, which is a DOE Office of Science User Facility supported under Contract DE-AC02-06CH11357.  CFG was supported in part by the IBM Einstein Fellow Fund at the Institute for Advanced Study, and also in part by grant NSF PHY-2309135 and the Gordon and Betty Moore Foundation Grant No. 2919.02 to the Kavli Institute for Theoretical Physics (KITP).}

\bibliography{references}

\end{document}